\documentclass[reprint,9pt,amsmath,amssymb,aps,superscriptaddress,longbibliography
]{revtex4-2}

\usepackage{extarrows}
\usepackage{graphicx}
\usepackage{dcolumn}
\usepackage{bm}
\usepackage[dvipsnames]{xcolor}
\usepackage[]{natbib}
\usepackage{url}

\usepackage{soul}

\newcommand{\eref}[1]{Eq.\,(\ref{#1})}
\newcommand{\eg}{\emph{e.g.}}
\newcommand{\ie}{\emph{i.e.}}
\mathchardef\mhyphen="2D

\begin{document}
	\title{Phase transitions in growing groups: How cohesion can persist}

    \author{Enrico Maria Fenoaltea}\email{enrico.fenoaltea@unifr.ch}
    \affiliation{Physics Department, University of Fribourg, Chemin du Mus\'ee 3, 1700 Fribourg, Switzerland}

    \author{Fanyuan Meng}\email{fanyuan.meng@hotmail.com}
    \affiliation{Research Center for Complexity Sciences, Hangzhou Normal University, Hangzhou, Zhejiang 311121, China}

    \author{Run-Ran Liu}
    \affiliation{Research Center for Complexity Sciences, Hangzhou Normal University, Hangzhou, Zhejiang 311121, China}

    \author{Matúš Medo}
    \affiliation{Department of Radiation Oncology, Inselspital, University Hospital of Bern, and University of Bern, 3010 Bern, Switzerland}
    \affiliation{Physics Department, University of Fribourg, Chemin du Mus\'ee 3, 1700 Fribourg, Switzerland}


\begin{abstract}
The cohesion of a social group is the group's tendency to remain united. It has important implications for the stability and survival of social organizations, such as political parties, research teams, or online groups. Empirical studies suggest that cohesion is affected by both the admission process of new members and the group size. Yet, a theoretical understanding of their interplay is still lacking.
To this end, we propose a model where a group grows by a noisy admission process of new members who can be of two different types. Cohesion is defined in this framework as the fraction of members of the same type and the noise in the admission process represents the level of randomness in the evaluation of new candidates. The model can reproduce the empirically reported decrease of cohesion with the group size. When the admission of new candidates involves the decision of only one group member, the group growth causes a loss of cohesion even for infinitesimal levels of noise.
However, when admissions require a consensus of several group members, there is a critical noise level below which the growing group remains cohesive. The nature of the transition between the cohesive and non-cohesive phases depends on the model parameters and forms 
a rich structure reminiscent of critical phenomena in ferromagnetic materials.
\end{abstract}
    

\maketitle

\section{Introduction}
A social group is an organized collection of individuals who share common interests or goals~\cite{stangor2015social}. Social groups are the building blocks of our society: they emerge naturally in larger social networks~\cite{newman2003social,jackson2010social,sawyer2005social}
and play an important role in opinion spreading \cite{galam2002minority, ecker2022psychological, wu2004information}, cultural diffusion \cite{henrich2004demography, centola2007homophily}, polarization \cite{baumann2020modeling}, and political impact \cite{denton2016social, garimella2018political}, among many others.
The formation and growth of social groups have been studied in both offline \cite{aronson1959effect, mcgrath1984groups,isakov2019structure, guimera2005team} and online~\cite{backstrom2006group,leskovec2008microscopic,lazer2009life,phan2015natural} environments.


In the context of complex networks, research
has focused on how
distinct growth mechanisms shape
the group's network structure \cite{krapivsky2001degree, jin2001structure, stadtfeld2020emergence} 
and what are the consequences of the former on various dynamical processes that can take place on social networks \cite{boccaletti2006complex,dorogovtsev2003evolution, schweitzer2022social, newman2001clustering}. Here, instead, we study how group growth affects cohesion. Defined as ``the tendency of a group to stick together and remain united''~\cite{carron2002team}, cohesion is crucial for the survival of a group \cite{pahl1991search, bruhn2009concept}, and hence must be considered before any social structure. Cohesion is a much-studied subject in psychology and sociology (see~\cite{fonseca2019social} for a recent review) with important implications. For example, cohesive groups have been shown to perform better in their tasks~\cite{carron2002team,casey2009sticking}. Among the main factors that influence cohesion is the admission process of new members \cite{aronson1959effect,gerard1966effects}. At the same time, empirical studies show that groups with more members tend to be less cohesive~\cite{delhey2007enlargements,wheelan2009group}. A successful group that attracts new members is thus in danger of gradually losing its cohesion, worsening its performance, and possibly even fragmenting into smaller cohesive groups. One example of well-documented group fragmentation due to a lack of group cohesion is the classical Zachary karate club~\cite{zachary1977information}, where a disagreement over lesson fees evolved into an ideological conflict among club members and finally resulted in the club's division. Another situation where a lack of cohesion can have serious consequences on society is that of political parties and governments. Indeed, mechanisms that favour the formation of heterogeneous coalition governments, \ie,  non-cohesive governments, tend to result in a high degree of fragmentation and instability, with negative social and economic consequences \cite{taylor1971party}.  
To avoid the above-mentioned detrimental outcomes, it is essential to understand the interplay between the admission process, group size, and group cohesion. Yet, to the best of our knowledge, no such integrative model exists.

Motivated by this lack of knowledge, we formulate a minimalist model for the evolution of a social group. In the model, a small homogeneous group gradually grows by adding new members through an admission process. 
We assume candidate members of two types: fit for the group (sharing goals or values with the founder members) or unfit for the group otherwise. Candidates are admitted or rejected based on positive or negative evaluations received from the group members.
Since member similarity contributes significantly to group cohesion \cite{hogg1993group,lott1965group, pham2022empirical}, in our framework we define cohesion as the fraction of fit group members.
More complicated definitions capturing other aspects of group cohesion~\cite{mcleod2013towards,carron2000cohesion}, such as the group's ability to work toward a given task, can be studied in the future. 
We assume that individual evaluations of candidates are at odds with homophily (a group member appreciating a candidate of the same type) with some small probability that we refer to as the evaluation noise. In the real world, the presence of noise is a joint result of carelessness and the intrinsic difficulty of the evaluation process. 
A careless interviewer, for example, has a high probability of wrongly assessing whether a job candidate is suitable or not, yet even the most meticulous interviewer has a non-zero chance of making a mistake~\cite{fernandez2006behavioral, porter2010dangerous}.
In addition to noise, the rigorousness of the admissions process in our framework is determined by how group members who evaluate a candidate are chosen, their number, and how individual evaluations are aggregated. Regarding the latter point, we consider for simplicity an aggregation based on unanimous consensus, meaning that all members involved in the evaluation must agree on admissions.

Our goal is to study the long-term effects of various admission processes on group cohesion in the presence of noise.
We find that if each candidate is evaluated by one group member, group cohesion markedly decreases with the group size. 
The outcome of the admission process is fundamentally different when admissions require a consensus of two or more group members. We find analytically that a critical evaluation noise level exists, above which group cohesion approaches the cohesion of a random group or is even lower. As the number of members involved in the admission process increases, the critical noise level increases. We show that at the critical point, the model exhibits a phase transition whose properties depend on the relative proportion of fit and unfit candidates who want to join the group. In particular, when fit and unfit candidates are equally probable, the phase transition belongs to the mean-field universality class. 

\section{Group growth Model}
The model is defined as follows. Consider individuals of two different types, $+1$ (fit for the group) and $-1$ (unfit for the group), respectively. Denote the type of group member $i$ by $\sigma_i$. The group initially consists of $N_0$ founder members that are all of type $+1$: $\sigma_i = +1$ for $i=1,\dots,N_0$. 
The growth of the group proceeds in discrete steps. In each step, one candidate member is considered to be admitted or rejected. The type of each candidate is drawn at random with equal probability (we relax this
assumption later). 
Denote the type of candidate in step $j$ as $c_j$. We assume that group members tend to positively evaluate candidates of the same type. Specifically, if group member $i$ is asked to evaluate candidate $j$ and $c_j \sigma_i = +1$, the evaluation is positive with probability $1-\eta$ and negative otherwise. If $c_j \sigma_i = -1$, the evaluation of candidate $j$ by member $i$ is positive with probability $\eta$ and negative otherwise. Here, $\eta\in[0,0.5]$ is the noise parameter that characterizes the level of randomness in the evaluation of candidates.

We assume that each candidate is evaluated by $m$ group members ($N_0\geq m$ to ensure a sufficient number of evaluators). A candidate is admitted only if all evaluations are positive. Within this framework, we consider two different ways to choose the evaluating group members: (1) uniformly (uniform case, UC) and (2) proportionally to the number of admissions to which the member has already contributed (preferential attachment, PA). The PA case mimics the accumulation of social capital by influential group members who thus evaluate more group members and leads to a scale-free distribution of social capital in the group~\cite{barabasi1999mean}. 
As an additional benchmark, we consider the case where one founder member evaluates each candidate (dictatorship, DS).

The growth of the group continues until a given group size, $N$, is reached. We then evaluate the resulting group cohesion, $C$, as the fraction of fit group members,
\begin{equation}
\label{cohesion}
C = \big\lvert\{i:\,\sigma_i=+1\}_{i=1}^N\big\rvert / N.
\end{equation}
A group composed solely of fit members has the cohesion of one (note that such a group can still be diverse when the members differ in features other than being fit for the given group).
When $N=N_0$, $C=1$ as all founder members are fit for the group. When $\eta=0$, unfit candidates always receive negative evaluations from fit group members. As the group initially consists of only fit members, all unfit candidates are rejected and the group cohesion remains one for any $N$. When $\eta=0.5$, the evaluations are not informative of the types of candidates and thus the probability of admitting an unfit candidate is $1/2$. As $N$ grows, group cohesion then approaches $1/2$ which is the cohesion level of a random group.

\section{Results}
\subsection{One evaluating member}
Let us first consider the case where the candidate is evaluated by only \emph{one} group member. To study cohesion analytically, we count the time in the growing process by the number of admitted members. In this way, the $N_0$ founder members are in the group at time $t=0$, while the $N$-th member is admitted at time $t=N-N_0$. We introduce the probability $P(t)$ that the $t$-th admitted member is fit for the group. As the group is assumed to initially consist of $N_0$ fit members, $P(0)=1$. As fit and unfit candidates are equally likely, if member $i$ admits a candidate, the candidate is fit with the probability
\begin{equation}
\label{eq:transition}
W(i)= P(i)(1-\eta)+[1-P(i)]\eta,
\end{equation}
where the two terms correspond to $\sigma_i=+1$ ($i$ is fit) and $\sigma_i=-1$ ($i$ is unfit), respectively. For the dictatorship case, only the founder members are allowed to evaluate, so $i=0$ and $P(t)=W(0)=1-\eta$. The cohesion is then given by 
\begin{equation}\label{DS1}
\overline{C(N,N_0,\eta)_{DS}}=(1-\eta)+ \frac{\eta N_0}{N}.
\end{equation}
As $N\to\infty$, group cohesion thus approaches $1-\eta$, i.e., it decreases linearly with noise.

\begin{figure*}
    \centering
    \includegraphics[scale=0.7]{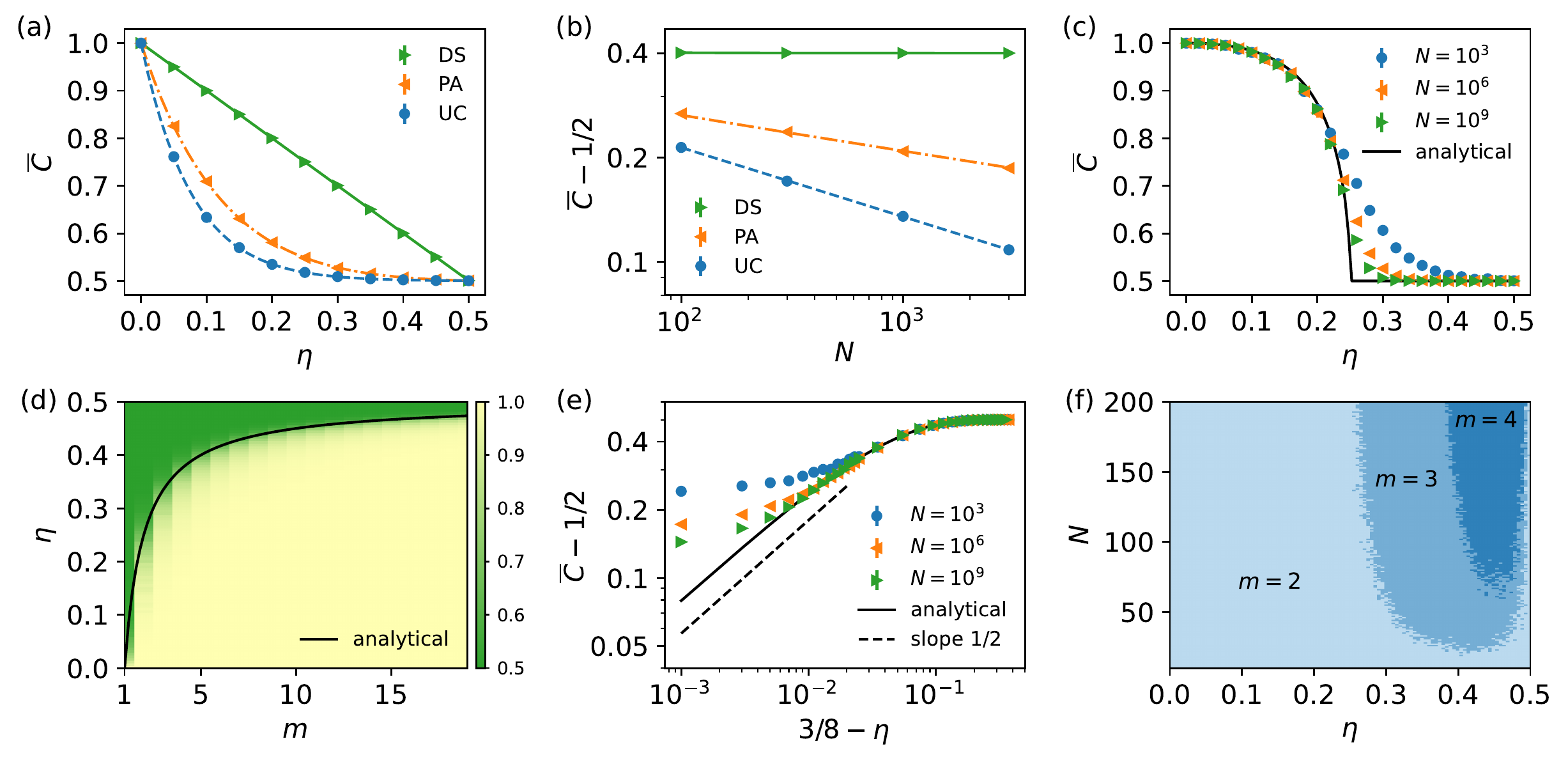}
    \caption{Analytical and numerical results when fit and unfit candidates are equally likely.
    (a) Mean cohesion, $\overline{C}$, for various ways (UC, PA, DS) of choosing one evaluating member vs.\ evaluation noise $\eta$ ($N_0=1$ and $N=10^3$).
    (b) Scaling of $\overline{C}-1/2$ with group size $N$ for various ways (UC, PA, DS) of choosing one evaluating member ($N_0=1$ and $\eta=0.1$).
    In panels (a) and (b), the solid lines represent the analytical solutions given by Eqs.\,(\ref{DS1}), (\ref{eq:N0_fixed_scaling}), and (\ref{eq:expected_cohesion_BA}). The symbols and their error bars (too small to be visible) represent means and standard errors based on 10\,000 model realizations.
    (c) Mean cohesion, $\overline{C}$, in the UC case with $m=2$ evaluating members vs.\ evaluation noise $\eta$ ($N_0=m$ and $N=10^3, 10^6, 10^9$). The solid line is the analytical limit when $N\to \infty$, given by \eref{2solu}.
    (d) The phase diagram of mean cohesion, $\overline{C}$, in the $(m,\eta)$ plane for the UC case ($N=10^6$ and $N_0=m$). The solid line is the analytical solution in the $N\to \infty$ limit, given by \eref{eta_c_general}.
    (e) Scaling of $\overline{C}-1/2$ in the UC case with $m=4$ evaluating members close to the critical noise level $\eta_c=3/8$ ($N_0=m$ and $N=10^3, 10^6, 10^9$). The solid line is the analytical limit when $N\to \infty$.
    (f) A phase diagram in the $(\eta,N)$ plane comparing multiple evaluators with a dictator. Each point is coloured by the smallest number of evaluators that are needed to outperform a single dictator (up to $m=4$).
    In panels from c--f, each point is an average over $500$ model realizations (error bars representing twice the SEM are too small to be visible).}
    \label{fig:first_big}
\end{figure*}

For the uniform case, the $t$-th admitted member can be evaluated by a random group member $i$ with $i<t$; $P(t)$ is hence obtained from $W(i)$ by averaging over all $i<t$ [$W(0)$ contributes $N_0$ times as there are $N_0$ founder members]. We get
\begin{equation}
\label{RCrecursive}
P(t+1)=\frac{N_0}{t+N_0}\,W(0)+\frac{1}{t+N_0}\,\sum_{i=1}^{t}W(i)
\end{equation}
which is a recursive equation for $P(t)$ with the initial condition $P(1)=1-\eta$ (as the first member is certainly evaluated by a founder member). \eref{RCrecursive} can be rearranged as
\begin{equation}
\label{solRC}
P(t+1)=\Big(1-\frac{2\eta}{N_0+t} \Big)P(t) + \frac{\eta}{N_0+t},
\end{equation}
which can be solved analytically. Averaging the solution over $t=0,\dots,N-N_0$ [again, the weight of $P(0)$ is $N_0$], the expected cohesion reads
\begin{equation}
\label{eq:C_anarchy}
\overline{C(N,N_0,\eta)_{UC}} = \frac{1}{2} \left(1+
\frac{\Gamma(N_0+1)\Gamma(N+1-2\eta)}{\Gamma(N_0+1-2\eta)\Gamma(N+1)} \right).
\end{equation}
When $\eta=0$, $\overline{C(N,N_0,\eta)}=1$ for any $N$ and $N_0$, as only fit candidates can be admitted. However, the expected cohesion decreases fast with $\eta$ and is always lower than that of the dictatorship case (Fig.~\ref{fig:first_big}a). When $N_0$ is fixed and $N\gg N_0$, \eref{eq:C_anarchy} implies
\begin{equation}
\label{eq:N0_fixed_scaling}
\overline{C(N,N_0,\eta)_{UC}} \approx \frac{1}{2} +
\frac{\Gamma(N_0+1)}{2\Gamma(N_0+1-2\eta)}\,N^{-2\eta}
\end{equation}
which approaches to $1/2$ as $N$ grows.
This means that, with only one evaluating member (randomly chosen), the group cohesion tends to the cohesion of a random group as the group grows, regardless of how small is the level of noise and how many are the founder members (Fig.~\ref{fig:first_big}b). This is in an agreement with the empirical studies showing that large groups are less cohesive~\cite{delhey2007enlargements,wheelan2009group}. In particular, the model with one evaluating member and uniform choice (UC) can be mapped on a recent opinion formation model~\cite{medo2021fragility} where a result analogous to \eref{eq:N0_fixed_scaling} has been derived using the master equation formalism.

To study the impact of introducing social capital in the model, we now consider choosing the evaluating member by the classical PA mechanism. To this end, we keep an activity counter, $k$, for all group members. This counter increases by $1$ for each participation in the admission of a new member. Without loss of generality, we set the initial activity counter to $a_0:=N_0-1$ and $1$ for all founder members and the later admitted group members, respectively (when $N_0=1$, we set $a_0=1$ to avoid zero counter of the sole founder member). The probability of choosing a member is assumed to be directly proportional to the activity counter. Using the continuum approximation~\cite{albert2002statistical} for the dynamics of the activity counter, we can write
\begin{equation}
\label{recursive_activity}
k_i(t+1) = k_i(t) + \frac{k_i(t)}{2t+N_0a_0},
\end{equation}
where the second term on the rhs reflects the PA selection mechanism. 

Analogously to \eref{RCrecursive}, the probability that the $t$-th admitted member is fit for the group, $P(t)$, for the PA case reads
\begin{equation}
\label{recursive_fitPA}
P(t+1) = N_0\frac{W(0)k_0(t)}{2t+N_0a_0}+
\sum_{i=1}^{t}\frac{W(i)k_i(t)}{2t+N_0a_0},
\end{equation}
where $W(i)$ is defined by \eref{eq:transition}. With the solution of \eref{recursive_activity} and the initial condition $P(1)=1-\eta$, \eref{recursive_fitPA} can be rearranged in a simpler form (similar to \eref{solRC})
\begin{equation}
P(t+1)=\left(1-\frac{2\eta}{2t+N_0a_0}\right)P(t)+
\frac{\eta}{2t+N_0a_0}.
\end{equation}
Similarly to the uniform case, the expected group cohesion is obtained by averaging the obtained $P(t)$ over $t=0,\dots,N-N_0$ (again assigning weight $N_0$ to $t=0$). We see (Fig.~\ref{fig:first_big}a) that the expected cohesion still decreases rapidly with $\eta$, yet it differs from the uniform case. When $N\to\infty$ and $N_0$ is fixed, the leading contribution to cohesion is
\begin{equation}
\label{eq:expected_cohesion_BA}
\overline{C(N,N_0,\eta)_{PA}} \approx \frac{1}{2} +
\frac{(1-2\eta)\Gamma(l_0+1)}{2(1-\eta)\Gamma(l_0+1-\eta)}\,N^{-\eta},
\end{equation}
where $l_0:=N_0(N_0-1)/2$. We find that group cohesion in the PA case is not only higher than in the uniform case (Fig.~\ref{fig:first_big}a), it also decays with $N$ slower (Fig.~\ref{fig:first_big}b): the scaling exponent is $\eta$ instead of $2\eta$. This higher robustness is due to preferential attachment effectively giving more power to early group members who, on average, are more likely to be fit than group members who join the group later. At the same time, the limit cohesion remains unchanged: as the group grows, its cohesion approaches the cohesion of a random group.


\subsection{More evaluating members}
We now study the model behaviour when $m>1$ members evaluate each candidate who is admitted only when all evaluations are positive. Because of the strong non-linearity of the process, we only treat the $N\to\infty$ limit.
Consider first the uniform case with $m=2$ and denote the evaluating members $i$ and $j$. For large $N$, the correlation between the types of $i$ and $j$ can be neglected, and the probabilities of admitting fit and unfit candidates can be factored as $W(i)W(j)$ and $[1-W(i)][1-W(j)]$, respectively, where $W(i)$ is given by \eref{eq:transition}.
As fit and unfit candidates are assumed to be equally likely, the probability that an admitted member is fit is
\begin{equation}
W_{ij}\equiv\frac{W(i)W(j)}{W(i)W(j)+[1-W(i)][1-W(j)]}.
\end{equation}
We build the solution again on the probability $P(t)$. As for \eref{RCrecursive}, $P(t)$ is obtained by averaging $W_{ij}$ over all possible pairs of evaluating members, each of whose is equally likely in the case of uniform selection. Thus, we can write
\begin{equation}\label{m=2_recursive}
P(t+1) = \frac{\binom{N_0}{2}W_{00}+N_0\sum_{i=1}^{t} W_{i0}+\sum_{j>i\ne 0}^{t} W_{ij}}{\binom{t+N_0}{2}},
\end{equation}
where the initial condition remains $P(0)=1$.
If $\lim\limits_{t\to\infty} P(t)$ exists, then
$\lim\limits_{N\to\infty} \overline{C(N)} = \lim\limits_{t\to\infty}P(t):=P$. To find the expected cohesion for $N\to\infty$, it thus suffices to obtain the stationary solution of \eref{m=2_recursive}. For large $t$, the main contribution to the numerator of \eref{m=2_recursive} comes from $W_{ij}$ where $i$ and $j$ are large. Denoting $W:=P(1-\eta)+(1-P)\eta$, we obtain
\begin{equation}\label{transition2}
P =\lim_{i,j\to\infty} W_{ij} = \frac{W^2}{W^2+(1-W)^2}.
\end{equation}
Besides the trivial solution $P=1/2$, this equation has two non-trivial solutions when $\eta<\eta_c=1/4$, representing the group composed of mostly fit and mostly unfit members, respectively. As we assume that the founder members are fit, the second solution is not physical, and we can write
\begin{equation}
\label{2solu}
P = \begin{cases}
\frac{1}{2}+\frac{\sqrt{1-4\eta}}{2(1-2\eta)} & \text{if $\eta<1/4$},\\
\frac{1}{2} & \text{if $\eta \geq 1/4$}.
\end{cases}
\end{equation}
As we said before, this $P$ is equal to the expected group cohesion in the limit $N\to\infty$. The expected cohesion thus undergoes a second-order phase transition at the critical noise $\eta_c=1/4$: from an ``ordered phase'', where most of the group's members are fit, to a ``disordered phase'', where the group is equally composed of fit and unfit members. Fig.~\ref{fig:first_big}c shows that the numerical simulations converge consistently to \eref{2solu} as $N$ increases, thus confirming our analytical results.

The above results can be generalized to $m>2$, leading to
\begin{equation}
\label{transitionm}
P=\frac{W^m}{W^m+(1-W)^m}.
\end{equation}
The solution again undergoes a second-order phase transition, this time at
\begin{equation}
\label{eta_c_general}
\eta_c=\frac{1}{2}-\frac{1}{2m}.
\end{equation}
as confirmed in Fig.~\ref{fig:first_big}d. This phase transition can be further characterized by studying its critical exponents. Expanding \eref{transitionm} around the critical value, we find $P - 1/2\propto (\eta_c-\eta)^\beta$, where $\beta=1/2$ for any $m\ge2$ (Fig.~\ref{fig:first_big}e). This shows that our model belongs to the mean-field universality class \cite{nishimori2010elements}. Note that \eref{transitionm} and \eref{eta_c_general} are in principle also valid for $m=1$, but in this case, we have a first-order phase transition at $\eta_c=0$ where the limit cohesion immediately drops from $1$ (for $\eta=0$) to $1/2$ (for $\eta>0$).

We find that more evaluating members dramatically improve the group's robustness to noise compared to only one evaluating member. This motivates us to compare with the dictatorship case where the limit cohesion is also above $1/2$.
Our results show that while cohesion for $m=1$ is always lower than for the dictatorship case, a noise range exists for $m>1$ where cohesion is higher than for the dictatorship case. Fig.~\ref{fig:first_big}f shows the regions of the parameter space $(\eta,N)$ 
where the unanimous decision of at least $m$ group members (up to $m=4$) is needed to have greater cohesion than in the dictatorship case.
Note that this region shrinks rapidly as $m$ increases. 
We thus see that with respect to group cohesion, relying on the consensus of several evaluating members is better than relying on a single dictator.

It is straightforward to show that when the choice of the $m$ evaluating members is driven by preferential attachment, the expected cohesion in the limit $N\to \infty$ coincides with that of the uniform case derived above. However, the PA mechanism makes convergence to the solution of \eref{transitionm} slower, as the weight of the founder members in \eref{m=2_recursive} is larger. 
This is analogous to \eref{eq:expected_cohesion_BA} converging to $1/2$ slower than \eref{eq:N0_fixed_scaling} for $m=1$.


\subsection{Less (or more) fit candidates}
So far we have assumed that fit and unfit candidates are equally likely. In the real world, however, there can be a marked asymmetry between the number of fit and unfit group's candidates. The fraction of fit candidates can be expected to be small, for example, for an attractive company or a top-rated university. 

\begin{figure*}
    \centering
    \includegraphics[scale=0.7]{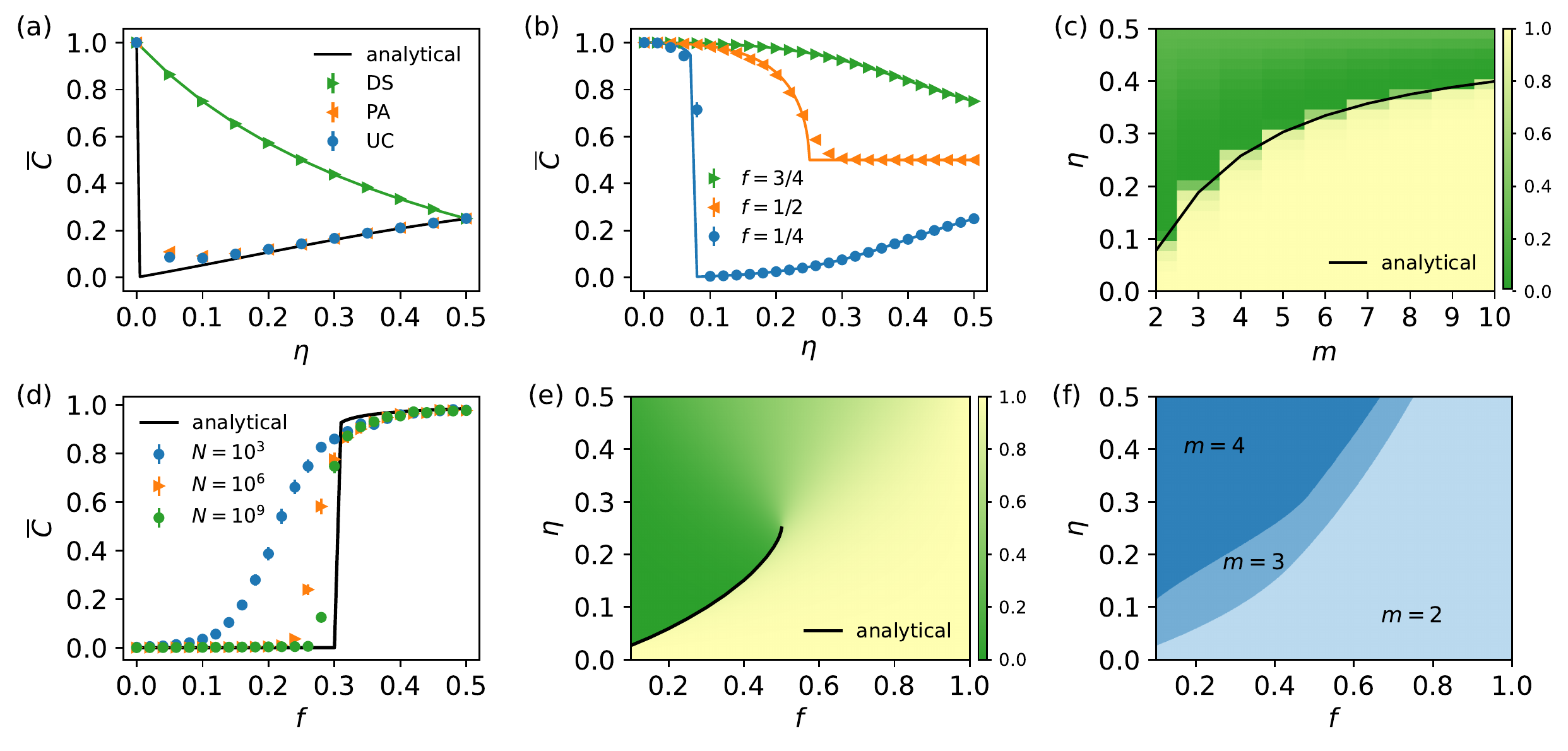}
    \caption{Analytical and numerical results when fit and unfit candidates are not equally likely ($f\ne1/2$).
    (a) Mean cohesion, $\overline{C}$, for various ways (UC, PA, DS) of choosing one evaluating member vs. evaluation noise $\eta$ ($N_0=1$, $N=10^3$, and $f=1/4$).
    (b) Mean cohesion, $\overline{C}$, in the UC case with $m=2$ evaluating members vs. evaluation noise $\eta$ ($N_0=m$, $N=10^9$, and $f=1/4, 1/2, 3/4$).
    (c) The phase diagram of mean cohesion, $\overline{C}$, in the $(m,\eta)$ plane for the UC case ($f=1/4$, $N_0=m$, and $N=10^6$).
    (d) Mean cohesion, $\overline{C}$, in the UC case with $m=2$ evaluating member vs. fraction of fit candidates $f$ ($\eta=0.1<\eta_c=1/4$, $N_0=m$, and $N=10^3, 10^6, 10^9$).
    (e) The phase diagram of mean cohesion, $\overline{C}$, in the $(f,\eta)$ plane for the UC case ($N=10^6$ and $N_0=m=2$).
    (f) A phase diagram in the $(f,\eta)$ space, comparing multiple evaluators with a dictator. Each point is colored by the smallest number of evaluators that are needed to outperform a single dictator (up to $m=4$). 
    In all panels, each data point is an average over $500$ model realizations (error bars representing twice the SEM are mostly too small to be visible). The solid lines represent numerical solutions of \eref{transition_f} which is valid for $N\to \infty$.}
    \label{fig:second_big}
\end{figure*}

To account for this, we introduce a new parameter, $f \in [0,1]$, which represents the (prior) probability
that
a candidate is fit for the group. When $f\ne 1/2$, obtaining an exact solution for group cohesion is difficult even when $m=1$. We thus treat only the case $N\to \infty$. In this limit, the steps leading to \eref{transitionm} can be repeated, this time explicitly
reflecting the probability $f$ that a candidate is fit when computing the probability that an admitted member is fit.
In this limit, the UC and PA cases have the same behaviour for the reasons discussed at the end of the previous section. In particular, \eref{transitionm} becomes
\begin{equation}
\label{transition_f}
P=\frac{fW^m}{fW^m+(1-f)(1-W)^m}
\end{equation}
which reduces to \eref{transitionm} when $f=1/2$. As before, this equation has two solutions, of which the higher one represents the studied case of fit founder nodes. 
Fig.~\ref{fig:second_big}a  and Fig.~\ref{fig:second_big}b shows the results for $m=1$ and $m=2$, respectively. 
When $\eta=1/2$, evaluations are uncorrelated with node types and admissions become random; the fraction of fit nodes in the group then approaches the fraction of fit nodes among the candidates and $\lim\limits_{N\to\infty}\overline{C} = f$.

When $f>1/2$, the solution of \eref{transition_f} is a continuous monotonically decreasing function of $\eta$. When $f<1/2$, we observe a first-order phase transition at a critical noise level $\eta_c$ that depends on $m$ and $f$ (Fig.~\ref{fig:second_big}c). At $\eta=\eta_c$, the limit expected cohesion drops from a value close to 1 to a value less than $1/2$. The size of the jump reduces to zero for $f=1/2$, where the nature of the phase transition changes from first to second order. 
Above $\eta_c$, cohesion grows until it reaches $\overline{C}=f$ in $\eta=1/2$. In the extreme case of $m=1$, we have $\eta_c=0$ for any $f<1/2$, so cohesion always grows with noise (excluding the point $\eta=0$), as shown in Fig.~\ref{fig:second_big}a. The increase of mean cohesion with $\eta$ for $\eta>\eta_c$ can be explained as follows. When $f<1/2$ and $\eta>\eta_c$, the high number of unfit candidates causes them to be gradually admitted more often than fit candidates due to evaluation errors. Once this happens, a group mostly composed of unfit members emerges and cohesion drops below $1/2$. Since this occurs already in the early stages of group formation, the system behaves as if there were unfit founder members and a majority of unfit candidates, which is the same (up to replacing $C$ with $1-C$) as fit founder members and a majority of fit candidates (cf. the results for $f=1/4$ and $f=3/4$ at $\eta>\eta_c$ in Fig.~\ref{fig:second_big}b). Thus, for any $f<1/2$ and $\eta>\eta_c$, the group reverses its initial composition.

The mathematical properties of the studied model allow us to interpret $f$ from a physicist's perspective. 
The model has no critical behaviour for $f>1/2$ and exhibits a second-order phase transition at $f=1/2$, similarly to what the infinite-range Ising model does when the external magnetic field is reduced to zero~\cite{domb2000phase}.
In this sense, $f$ is analogous to the external magnetic field and $f=1/2$ corresponds to zero field. Cohesion and the evaluation noise are, in turn, analogous to magnetization and temperature in the Ising model, respectively. When $f<1/2$, a first-order phase transition emerges in our model that is absent in the standard Ising model.
This is due to the symmetry breaking of the system caused by the initial conditions, \ie, the type of founder nodes. When $\eta<\eta_c$, cohesion is larger than $1/2$ even when $f<1/2$, analogous to a material initially magnetized in one direction which, for temperature not too high, remains magnetized in the same direction even if the external magnetic field points in the opposite direction. If we fix evaluation noise at $\eta<\eta_c(f=1/2)$ and vary $f$, cohesion undergoes a first-order phase transition at some critical value $f_c$ which depends on $\eta$ and $m$ (see Fig.~\ref{fig:second_big}d and Fig.~\ref{fig:second_big}e).

Finally, we study the dictatorship case when $f\ne1/2$. It is straightforward to show that in this case cohesion is no longer linear in $\eta$ but, for $N \to \infty$, it is given by $f(1-\eta)/(f-2f\eta+\eta)$, 
as shown in Fig.~\ref{fig:second_big}a. We compare this result with the solution of \eref{transition_f} in the $(f,\eta)$-phase diagram of Fig.~\ref{fig:second_big}f. Again, the area in this parameter space where at least $m$ evaluating members are needed to outperform the dictator decreases with $m$. In particular, when $m$ is sufficiently large, there exists a value of $f>1/2$ such that $m$ evaluating members are better than the dictator for any $\eta$. These findings again suggest that, when external conditions (\ie, noise and the fraction of unfit candidates) are not too adverse, it is preferable to rely on the consensus of several evaluating members rather than the judgments of a single dictator.

\section{Conclusion and Discussion}
We introduced a simple, yet not trivial, model of group formation to study the dynamics of group cohesion. We show that the number of members involved in the evaluation of new candidates is crucial to determine cohesion in large groups. As the level of randomness in the admission process increases, the system undergoes a phase transition.
Above a critical noise level, large groups cannot remain cohesive. However, the more members evaluate each candidate, the higher the critical noise level. In the extreme case of only one evaluating member, the critical noise is zero. Growing groups then gradually lose their cohesion regardless of how small the evaluation noise is. This not only agrees with empirical studies on the relation between group size and cohesion\cite{delhey2007enlargements,wheelan2009group} but also suggests an alternative and complementary mechanism for the observed group fragmentation in the society \cite{pham2021balance, minh2020effect}. This fragmentation is observed in groups that form spontaneously, corresponding to the case $m=1$ in our model where a single group member can decide (see \cite{backstrom2006group} for an analysis of groups of friends and online communities without an established admission process). More formal groups where the admission of new applicants follows a rigorous process can be modelled by increasing the number of evaluating group members.

We further considered the case where one fixed group member (dictator) decides all admissions. Although the dictator always performs better than one evaluator chosen at random, requiring a consensus on new admissions between several members is a better strategy if the evaluation noise is not too high.
Finally, we investigated the situation where the fractions of fit and unfit candidates are different. 
When unfit candidates are more frequent than fit candidates, the system undergoes a discontinuous transition at a critical noise above which the group becomes mostly composed of unfit members. 
Note that such a group is in principle also cohesive. However, this is still not a desirable outcome, as the unfit members may be unable to fulfill the group's original purpose. The unfit majority is furthermore in conflict with the fit founding members which 
can cause tensions and eventually lead to the group's disintegration. This transition can correspond to, for example, a significant party policy shift. An example of this is the 5-star movement, an Italian political party that, in the course of its evolution and growth, changed from being an eurosceptic party to a pro-European one \cite{5star}. This is an interesting feature of our model and we argue that it could advance the understanding of group opinion change phenomena, such as parties' position shifts (several examples of this are analyzed, for instance, in \cite{adams2012causes, haupt2010parties}).

In reality, the factors that contribute to the loss of cohesion of a growing group are multifaceted and not easily identifiable (\eg, psychological factors). Nonetheless, our model focuses on the most-essential group growth mechanism and suggests that group size, randomness, and the admission processes jointly affect group cohesion in a non-trivial way.

Among possible extensions of our work are different admission rules (\eg, majority voting) and the introduction of node-type dynamics as a result of peer influence. An example of how the latter can be implemented in our model comes from \cite{galam1991towards}, a pioneering paper that has inspired many subsequent models in the field of social physics \cite{castellano2009statistical}. In this paper, the authors introduce a non-ergodic theory of social collective phenomena, postulating that the dynamics of opinions in a system of interacting individuals evolves toward a configuration that maximizes the sum of conformity (analogous to cohesion in our model) and group entropy (representing the potential for innovation given by the number of possible configurations with a certain value of conformity). Similar to our findings, they showed that, by varying what they call the divergence parameter (analogous to noise in our model), there is a threshold value below which the group becomes biased toward an opinion (in our language, cohesion is positive). They also show that the larger the group size, the less the tendency to group polarization. However, here (and in the other subsequent works) the authors do not consider how the group grows, but only what are its equilibrium properties given its size. Hence, by integrating these well-established results with our group growth mechanism based on the admission process of new candidates, we believe our work can be further enriched and set a basis for a new research direction in the interdisciplinary field of social physics.

From a different perspective, fit and unfit nodes can be interpreted as individuals with different characteristics, while the evaluation noise then represents a tolerance to diversity. Our approach can then be used to study the relationship between group size, cohesion, and diversity. In addition, one can introduce the concept of the growth cost given by the time and total effort required to form a group with the desired size. Finally, as our model is closely related to a recent model of opinion formation~\cite{medo2021fragility}, our results can show how to prevent the formation of unreliable opinions or erroneous inference from complex network data~\cite{fajardo2022node}.

\vspace*{12pt}
We thank J.L. Juul, M.S. Mariani, and M.A. Schwarz who improved this paper with relevant comments. This work was partially supported by the Swiss National Science Foundation (grant no. $200020\_182498/1$). F.M. is supported by the China Scholarship Council.

\vspace*{12pt}
E.M.F. and F.M. performed calculations, wrote, and conceptualized this work; M.M. wrote and conceptualized this work; R.L. performed calculations. E.M.F. and F.M. equally contributed to this paper.

\bibliography{general}
\end{document}